\documentclass[final,3p,times,10pt]{elsarticle}

\usepackage{multirow,setspace,times,amssymb,amsmath,graphicx,color,rotating,subfigure,url}
\usepackage{lineno}
\usepackage{natbib}
\usepackage{mathrsfs}
\usepackage{makecell}
\journal{Elsevier}

\begin{document}
%\begin{CJK*}{GBK}{Song} % Use default fonts from CJK (see below)

\begin{frontmatter}

\title{Tetradic motif profiles of horizontal visibility graphs}

\author[RCE,SB]{Wen-Jie Xie}
\author[RCE,DM]{Rui-Qi Han}
\author[RCE,SB,DM]{Wei-Xing Zhou\corref{WXZ}}
\cortext[WXZ]{Corresponding to: 130 Meilong Road, P.O. Box 114, School of Business, East China University of Science and Technology, Shanghai 200237, China.}
\ead{wxzhou@ecust.edu.cn}

\address[RCE]{Research Center for Econophysics, East China University of Science and Technology, Shanghai 200237, China}
\address[SB]{Department of Finance, East China University of Science and Technology, Shanghai 200237, China}
\address[DM]{Department of Mathematics, East China University of Science and Technology, Shanghai 200237, China}

\begin{abstract}
  Network motif analysis is a useful tool for the investigation of complex networks. We study the profiles of tetradic motifs in horizontal visibility graphs (HVGs) converted from multifractal binomial measures, fractional Gaussian noises, and heartbeat rates. The profiles of tetradic motifs contains the spatial information (visibility) and temporal information (relative magnitude) among the data points in the corresponding time series. For multifractal binomial measures, the occurrence frequencies of the tetradic motifs are determined, which converge to a constant vector $(2/3,0,8/99,8/33,1/99,0)$. For fractional Gaussian noises, the motif occurrence frequencies are found to depend nonlinearly on the Hurst exponent and the length of time series. These findings suggest the potential ability of tetradic motif profiles in distinguishing different types of time series. Finally, we apply the tetradic motif analysis to heartbeat rates of healthy subjects, congestive heart failure (CHF) subjects, and atrial fibrillation (AF) subjects. Different subjects can be distinguished from the occurrence frequencies of tetradic motifs.
\end{abstract}


\end{frontmatter}

\section{Introduction}
\label{S1:Introduction}

Time series motif analysis has been widely used in diverse fields
\cite{Mueen-Keogh-Zhu-Cash-Westover-2009-SIAM,Chiu-Keogh-Lonardi-ACM,Mueen-Keogh-2010-KDD,Mcgovern-Rosendahl-Brown-Droegemeier-2011-DMKD,Gomes-Batista-2015-ASTL,Wang-Chng-Li-2010-PRL,Son-Anh-2016-KIS,Xie-Han-Zhou-2019-EPL}.
In recent years, there have been many methods to transform time series into complex networks. The problem of time series analysis has also been transformed to complex network analysis
 \cite{Zou-Pazo-Romano-Thiel-Kurths-2007-PRE,Marwan-Romano-Thiel-Kurths-2007-PR,Zhang-Small-2006-PRL,Xu-Zhang-Small-2008-PNAS,Lacasa-Luque-Ballesteros-Luque-Nuno-2008-PNAS,Marwan-2008-EPJST,Marwan-Donges-Zou-Donner-Kurths-2009-PLA,Donner-Zou-Donges-Marwan-Kurths-2010-NJP,Gao-Small-Kurths-2016-EPL}.
Among these, the visibility graph (VG) \cite{Lacasa-Luque-Ballesteros-Luque-Nuno-2008-PNAS,Lacasa-Luque-Luque-Nuno-2009-EPL,Ni-Jiang-Zhou-2009-PLA,Qian-Jiang-Zhou-2010-JPA,Yang-Wang-Yang-Mang-2009-PA,Elsner-Jagger-Fogarty-2009-GRL}
and horizontal visibility graphs (HVGs) \cite{Lacasa-Luque-Luque-Nuno-2009-EPL,Luque-Lacasa-Ballesteros-Luque-2009-PRE} have attracted much attention. Many researchers have developed many variants of VGs and HVGs, such as Markov-binary visibility graph (MBVG) \cite{Ahadpour-Sadra-ArastehFard-2014-IS}, limited penetrable visibility graph \cite{Zhou-Jin-Gao-Luo-2012-APS,Gao-Cai-Yang-Dang-Zhang-2016-SR,Gao-Cai-Yang-Dang-2017-PA},  parametric natural visibility graph (PNVG) \cite{Bezsudnov-Snarskii-2014-PA,Snarskii-Bezsudnov-2016-PRE}, binary visibility graph (BVG) \cite{Ahadpour-Sadra-2012-IS}, multiplex horizontal visibility graph \cite{Zou-Donner-Marwan-Small-Kurths-2014-NPG,Lacasa-Nicosia-Latora-2015-SR,Bianchi-Livi-Alippi-Jenssen-2017-SR}.

With the mapping algorithms mentioned above, the problem of time series motif analysis has also been transformed to complex network motif analysis. Network motifs are regarded as recurrent and statistically significant sub-graphs or patterns \cite{Milo-ShenOrr-Itzkovitz-Kashtan-Chklovskii-Alon-2002-Science,Milo-Itzkovitz-Kashtan-Levitt-ShenOrr-Ayzenshtat-Sheffer-Alon-2004-Science}, which reflect the microstructure characteristics of networks.
Their occurrence patterns can be used to define similarity of networks \cite{Milo-ShenOrr-Itzkovitz-Kashtan-Chklovskii-Alon-2002-Science,Milo-Itzkovitz-Kashtan-Levitt-ShenOrr-Ayzenshtat-Sheffer-Alon-2004-Science}.
Iacovacci and Lacasa defined the sequential HVG motifs \cite{Iacovacci-Lacasa-2016a-PRE} and the sequential VG motifs \cite{Iacovacci-Lacasa-2016b-PRE}, which can distinguish different dynamics of time series. They extracted the motif profiles from experimental heart-rate series to perform unsupervised learning and disentangle meditative from other relaxation states. These studies stimulate us to investigate the profiles of ``natural'' tetradic motifs in HVGs, which advances previous results on tetradic HVG motifs of fractional Brownian motions \cite{Xie-Zhou-2011-PA}.

In this paper, we construct horizontal visibility graphs from multifractal binomial measures, fractional Gaussian noises and heartbeat rates.
We investigate the profiles of tetradic motifs for multifractal binomial measures and simulate fractional Gaussian noises. The impacts of time series lengths and correlation strengths on motif occurrence frequencies are studied numerically. These findings imply that different types of time series may have different profiles of tetradic HVG motifs, which can be used to distinguish different time series.
We finally apply the method to analyze the heartbeat rates of healthy subjects, congestive heart failure (CHF) subjects, and atrial fibrillation (AF) subjects and distinguish different patients.

\section{Tetradic motifs}

We transform the time series $\{x_n\}_{n = 1}^{N}$ into a network $G = \langle V, E\rangle$ based on the horizontal visibility algorithm \cite{Luque-Lacasa-Ballesteros-Luque-2009-PRE}, where $V = \{v_n\}_{n = 1}^{N}$ is the set of nodes in $G$ corresponding to the data points in the time series $\{x_n\}_{n = 1}^{N}$ and
$E = \{e_{ij}\}$ is the adjacency matrix of $G$. If node $v_{i}$ and node $v_{j}$ are horizontally visible to each other, an edge $e_{ij} = 1$ is drawn between them. It means that no data
points between node $v_{i}$ and $v_{j}$ are greater than the value of these two nodes. Mathematically, we have
\begin{equation}
   x_{i}, x_{j} >  x_{n}~~~~~~~~~~(i<n<j).
    \label{Eq:HVG}
\end{equation}

\begin{figure}[htb]
  \centering
\includegraphics[width=16cm]{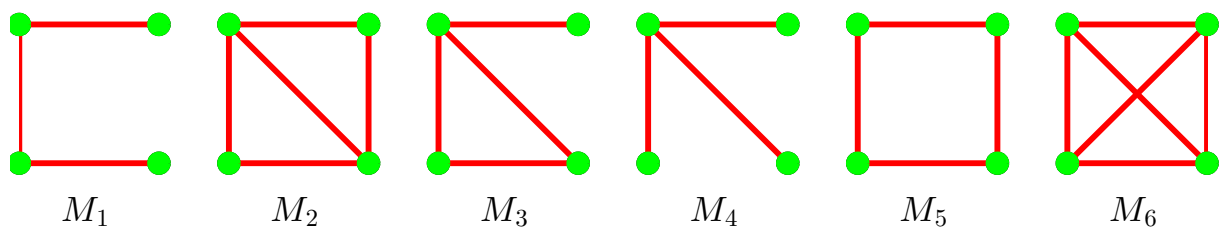}
  \caption{\label{Fig:TetradicMotifs:Def} Six undirected tetradic motifs.}
\end{figure}

The HVG algorithm converts the temporal features of a time series into the spatial characteristics of the corresponding HVG, such as network motifs. Here we investigate the undirected tetradic motifs of HVGs for understanding the characteristics of corresponding time series. Undirected tetradic motifs have six types, as shown in Fig.~\ref{Fig:TetradicMotifs:Def}. The six motifs are denoted as $M_1$, $M_2$, $M_3$, $M_4$, $M_5$ and $M_6$.

The occurrence count of motif $M_i$ in the HVG is denoted as $N_i$. The corresponding occurrence frequency is thus expressed as follows,
\begin{equation}
  f_{i} = {N_i}/{\sum_{i=1,\ldots,6} N_i}.
\end{equation}
Each motif in Fig.~\ref{Fig:TetradicMotifs:Def} represents the specific relationship between the data points. The occurrence frequency $f_{i}$ of motif $M_{i}$ reflects the characteristics of time series from the microscopic perspective. We will analyze the profiles of tetradic motifs for time series generated by simulations and in real world.

\section{Tetradic motif profiles for multifractal binomial measures}

We use the $p$-model \cite{Meneveau-Sreenivasan-1987-PRL} to generate multifractal binomial measures and to construct HVG for network motif analysis. The generation procedure of a multifractal binomial measure starts from a measure uniformly distributed on an interval $[0,1]$. At the first step, the measure is redistributed on the interval, $\mu_{1,1}=\mu p_1$ to the first half interval and $\mu_{1,2}=\mu p_2=\mu (1-p_1)$ to the second half interval. At the $(m+1)$-th step, the measure $\mu_{k,i}$ on each of the $2^m$ line segments is redistributed into two parts, where
\begin{equation}
 \mu_{k+1,2i-1}=\mu_{k,i}p_1
\end{equation}
and
\begin{equation}
 \mu_{k+1,2i}=\mu_{k,i}p_2.
\end{equation}
After the generation process is iterated over $m$ times, we get a time series with $2^m$ data points.

\begin{figure}[htb]
  \centering
  \includegraphics[width=8cm]{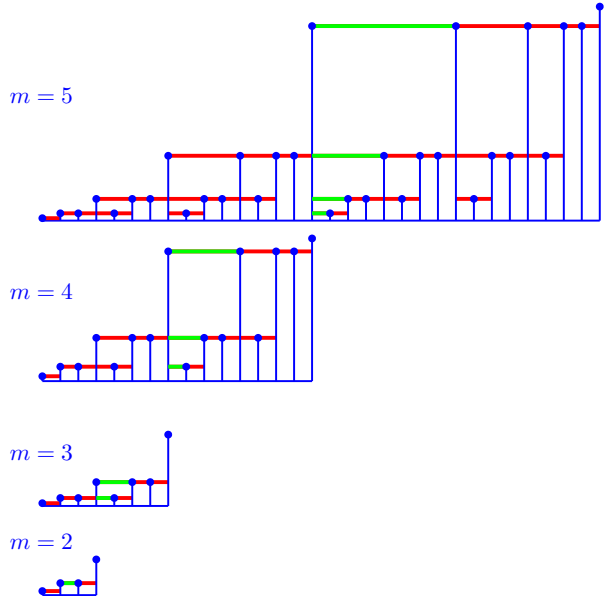}
  \caption{\label{Fig:HVGs:pmodel} (Color online.)  Illustrative examples showing the construction of HVGs from multifractal binomial measures.}
\end{figure}

In our analysis, the parameters $p_1$ of the $p$-model is 0.25 and $p_2$ is $1-p_1=0.75$. Obviously, other choices of $p_1$ do not affect the results because they will not change the visibility feature among different data points. Fig.~\ref{Fig:HVGs:pmodel} shows the structure of the HVGs mapped from the multifractal binomial measures with $m=2$, $m=3$, $m=4$ and $m=5$.
The time series of multifractal binomial measures with length $N=2^m$ is spliced from two time series with length $2^{m-1}$. Similarly, we construct the HVG with $2^{m}$ nodes by splicing two HVGs with $2^{m-1}$ nodes \cite{Xie-Han-Jiang-Wei-Zhou-2017-EPL}.

We denote the number of network motifs $M_i$ as $N_i(m)$ in the HVG corresponding to the multifractal binomial measures with length $N=2^m$.
Using motif recognition algorithm, we calculate the number $N_i(m)$ of network motifs $M_i$ corresponding to the multifractal binomial measures with length $N=2^m$. The results are shown in Table~\ref{Tb:pmodel:Nmi} for $m=1,2, \cdots, 8$. It is observed that $N_2=N_6=0$. Actually, $N_6=0$ holds for the HVGs mapped from any time series \cite{Xie-Zhou-2011-PA}. The observation of $N_2=0$ holds for other $m$ values, which can be reached by using $\eta_i(m)=0$ in Eq.~(\ref{Eq:pmodel_Skm}) for motif $M_2$.

\begin{table}[htp]
\centering
%\vspace{-10bp}\addtocounter{table}{-1}\renewcommand{\tablename}{Table}
\caption{The count $N_i(m)$ of network motif $M_i$ in the HVG mapped from multifractal binomial measures with $2^m$ data for $m=1,2,\cdots,8$.}
\medskip
\renewcommand\arraystretch{1}
\label{Tb:pmodel:Nmi}
%\begin{tabular}{ p{0.03\textwidth} p{0.06\textwidth} p{0.06\textwidth}p{0.06\textwidth}p{0.06\textwidth}p{0.06\textwidth}p{0.06\textwidth}}
\begin{tabular}{ccccccclll}
  \hline
   $m$ &$N_1$ & $N_2$ & $N_3$& $N_4$& $N_5$& $N_6$ \\
    \hline
1 & 0 & 0 & 0  & 0  & 0 & 0    \\
2 & 1 & 0 & 0  & 0  & 0 & 0    \\
3 & 6 & 0 & 2  & 0  & 0 & 0    \\
4 & 25 & 0 & 8  & 4  & 1 & 0    \\
5 & 89 & 0 & 22  & 22  & 3 & 0    \\
6 & 260 & 0 & 52  & 76  & 7 & 0    \\
7 & 666 & 0 & 114  & 212  & 15 & 0    \\
8 & 1567 & 0 & 240  & 524  & 31 & 0    \\
%9 & 3487 & 0 & 494  & 1202  & 63 & 0    \\
%10 & 7478 & 0 & 1004  & 2628  & 127 & 0    \\
%11 & 15648 & 0 & 2026  & 5568  & 255 & 0    \\
%12 & 32217 & 0 & 4072  & 11556  & 511 & 0    \\
%13 & 65629 & 0 & 8166  & 23662  & 1023 & 0    \\
%14 & 132776 & 0 & 16356  & 48028  & 2047 & 0    \\
%~~~~15~~~~ & ~~~~267446~~~~  & ~~0~~ &~~~~ 32738~~~~  & ~~~~96940~~~~   &~~~~ 4095~~~~ & ~~0~~    \\
\hline
\end{tabular}
\end{table}

According to previous works \cite{Xie-Zhou-2011-PA,Iacovacci-Lacasa-2016a-PRE,Iacovacci-Lacasa-2016b-PRE}, we found that $N_5 = 0$ for most HVGs. However, the occurrence frequencies of $M_5$ is not equal to zero for multifractal binomial measures. Motif $M_5$ exists if and only if $x_1, x_4>x_2=x_3$, then $e_{12}=e_{23}=e_{34}=e_{14}=1$. Motif $M_5$ rarely appears in HVGs mapped from a random time series, but the condition $x_2=x_3$ in multifractal binomial measures is easy to satisfy, as shown in Fig.~\ref{Fig:HVGs:pmodel}.
The multifractal binomial measures with $N=2^m$ numbers can be represented as $\{p_1^{k_n}(1-p_1)^{m-k_n}\}_{n = 1,...,N} $, where $k_n\in\{0,1,2,...,m\}$. This time series can be determined by the variable $k_n$, so the original time series can be expressed as $\{k_n\}_{n = 1,...,N}$, where $k_n$ can only be selected from the collection $\{0,1,2,...,m\}$. Only $m+1$ numerical values exist in the time series of multifractal binomial measures with length $2^m$, where the condition $x_2=x_3$ is easily satisfied and motif $M_5$ appears surely for multifractal binomial measures.

Furthermore, we will obtain the expression for the occurrence count $N_i(m)$ of motif $M_i$.
We put two HVGs containing $2^{m-1}$ nodes together to obtain an HVG with $2^m$ nodes. In these processes, the original number of network motifs is unchanged. After adding $m-1$ edges in the HVG, the network structure changes. To simplify the presentation, we use a polynomial function $\eta_i(m)$ to denote the number of new motif $M_i$. The count of network motif $M_i$ can be expressed as follows:
\begin{equation}
  N_i(m)=2N_i(m-1)+\eta_i(m),~~~~i=1,...,6.
  \label{Eq:pmodel_Skm}
\end{equation}
The polynomial function $\eta_i(m)$ differs for different motifs. We are not able to give an explicit mathematical expression for the polynomial function $\eta_i(m)$. Inspired by the derivation and theoretical results in the literature \cite{Xie-Han-Jiang-Wei-Zhou-2017-EPL}, we assume that the count $N_i(m)$ of motif $M_i$ satisfies the following mathematical expression:
\begin{equation}
  N_i(m)=a_i2^{m}+b_im^3+c_im^2+d_im+e_i,~~~~i=1,...,6.
  \label{Eq:pmodel:Nmi}
\end{equation}
The parameters $a_i$, $b_i$, $c_i$, $d_i$, and $e_i$ are the coefficients to be determined.

Substituting $m$ and $N_i(m)$ in Table~\ref{Tb:pmodel:Nmi} into Eq.~(\ref{Eq:pmodel:Nmi}), we can obtain the parameters $a_i$, $b_i$, $c_i$, $d_i$, and $e_i$, as well as the mathematical expression $N_i(m)$. For each motif $M_i$, we substituted $N_i(4) $, $N_i(5)$, $N_i(6)$, $N_i(7)$, and $N_i(8)$ into Eq.~(\ref{Eq:pmodel:Nmi}) to fit the parameters $a_i$, $b_i$, $c_i$, $d_i$, and $e_i$. By solving four six-element linear equations, we obtain the parameters $a_i$, $b_i$, $c_i$, $d_i$, and $e_i$ for the motifs, leading to the following expression:
\begin{equation}
    N_i(m)= \left\{
    \begin{aligned}
      &\frac{33}{4}2^{m}-\frac{2}{3} m^3-\frac{5}{2} m^2-\frac{29}{6} m-5, &i=1\\
      &0, &i=2\\
      &2^{m}-2 m, & i=3\\
      &3\times2^{m}-\frac{1}{3} m^3- m^2-\frac{2}{3} m-4, & i=4\\
      &\frac{1}{8}2^{m}-1, & i=5\\
      &0, &i=6\\
    \end{aligned}
    \right.
    \label{Eq:pmodel:Nm:all}
\end{equation}

The total count of tetradic motifs is $N(m)=\sum_{i=1,...,6}N_i(m)$, which can be expressed as follows:
\begin{equation}
  N(m)=\frac{99}{8}\times2^{m}- m^3-\frac{7}{2} m^2-\frac{45}{6} m-10
 \label{Eq:pmodel_Msum}
\end{equation}
We can see that the count of motifs increases exponentially with respect to the length of time series. The occurrence frequencies $f_i$ of the motifs are
\begin{equation}
 f_i(m)=\frac{N_i(m)}{N(m)}
%    f_i(m)= \\
%    \left\{
%    \begin{aligned}
%      &\frac{\frac{33}{4}2^{m}-\frac{2}{3} m^3-\frac{5}{2} m^2-\frac{29}{6} m-5}{\frac{99}{8}2^{m}- m^3-\frac{7}{2} m^2-\frac{45}{6} m-10}, & i=1\\
%      &0, & i=2\\
%      &\frac{2^{m}-2 m}{\frac{99}{8}2^{m}- m^3-\frac{7}{2} m^2-\frac{45}{6} m-10}, & i=3\\
%      &\frac{3\times2^{m}-\frac{1}{3} m^3- m^2-\frac{2}{3} m-4}{\frac{99}{8}2^{m}- m^3-\frac{7}{2} m^2-\frac{45}{6} m-10}, & i=4\\
%      &\frac{\frac{1}{8}2^{m}-1}{\frac{99}{8}2^{m}- m^3-\frac{7}{2} m^2-\frac{45}{6} m-10}, & i=5\\
%      &0, & i=6\\
%    \end{aligned}
%    \right.
    \label{Eq:pmodel:fm:all}
\end{equation}
As $m$ increases, the exponential part $a_i2^{m}$ in the formula $N_i(m)$ will be much larger than the polynomial part $b_im^3+c_im^2+d_im+e_i$. Hence, when $m\rightarrow\infty$, the polynomial part can be ignored, and the occurrence frequency of each motif will converge to a constant. We have
\begin{equation}
  %   f_i(m\rightarrow\infty)=\left\{
    \lim_{m\to\infty} {\mathbf{f}} = \lim_{m\to\infty} \{f_i(m\rightarrow \infty): i=1,2,\cdots, 6\}= \left(\frac{2}{3},0,\frac{8}{99},\frac{8}{33},\frac{1}{99},0\right)
     %\left\{
%    \begin{aligned}
%      &\frac{a_1}{a_1+a_3+a_4+a_5}=2/3, &i=1\\
%     &0, &i=2\\
%      &\frac{a_3}{a_1+a_3+a_4+a_5}=8/99, &i=3\\
%      &\frac{a_4}{a_1+a_3+a_4+a_5}=8/33, &i=4\\
%      &\frac{a_5}{a_1+a_3+a_4+a_5}=1/99, &i=5\\
%     &0, &i=6\\
%
%    \end{aligned}
%    \right.
    \label{Eq:Eq:pmodel_pkm_inf}
\end{equation}

\begin{figure}[t!]
  \centering
  \includegraphics[width=15cm]{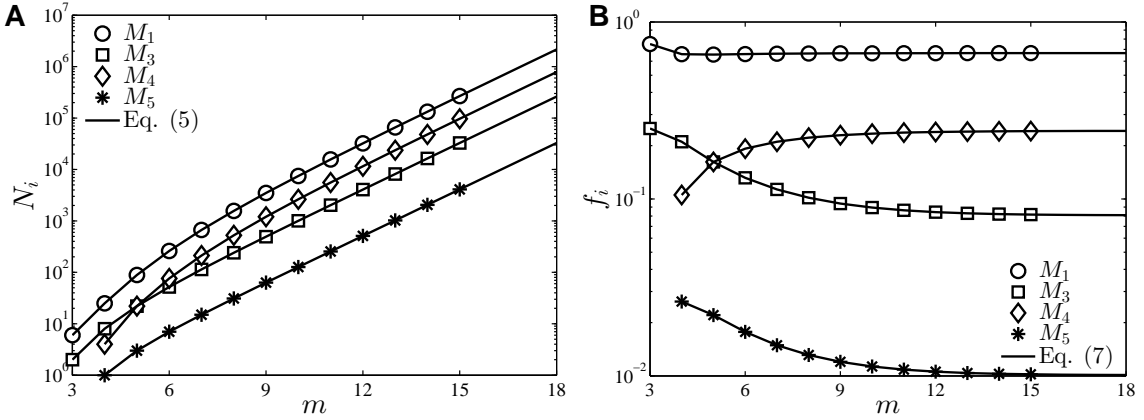}
  \caption{(A) Occurrence counts $N_i$ of motifs $M_i$ for $i=1,3,4,5$ in the HVGs mapped from multifractal binomial measures. The solid lines correspond to Eq.~(\ref{Eq:pmodel:Nm:all}). (B) Occurrence frequencies $f_{i}$ of motifs $M_i$. The solid lines correspond to Eq.~(\ref{Eq:pmodel:fm:all}).}
  \label{Fig:HVG:Mimic:pmodel}
\end{figure}

Note that Eq.~(\ref{Eq:pmodel:Nm:all}) and Eq.~(\ref{Eq:pmodel:fm:all}) are obtained using numerical results with $m\leq8$. We now check if these expressions hold for larger values of $m$. Fig.~\ref{Fig:HVG:Mimic:pmodel}(A) shows the occurrence counts $N_i$ of motifs $M_i$ for $i=1,3,4,5$ in the HVGs mapped from multifractal binomial measures of different lengths, where the corresponding Eq.~(\ref{Eq:pmodel:Nm:all}) are drawn as solid lines. Fig.~\ref{Fig:HVG:Mimic:pmodel}(B) presents the occurrence frequencies $f_{i}$ of motifs $M_i$, where the solid lines correspond to Eq.~(\ref{Eq:pmodel:fm:all}). It is found that the expressions hold for larger $m$ values.

\begin{figure}[t!]
\centering
\includegraphics[width=15cm]{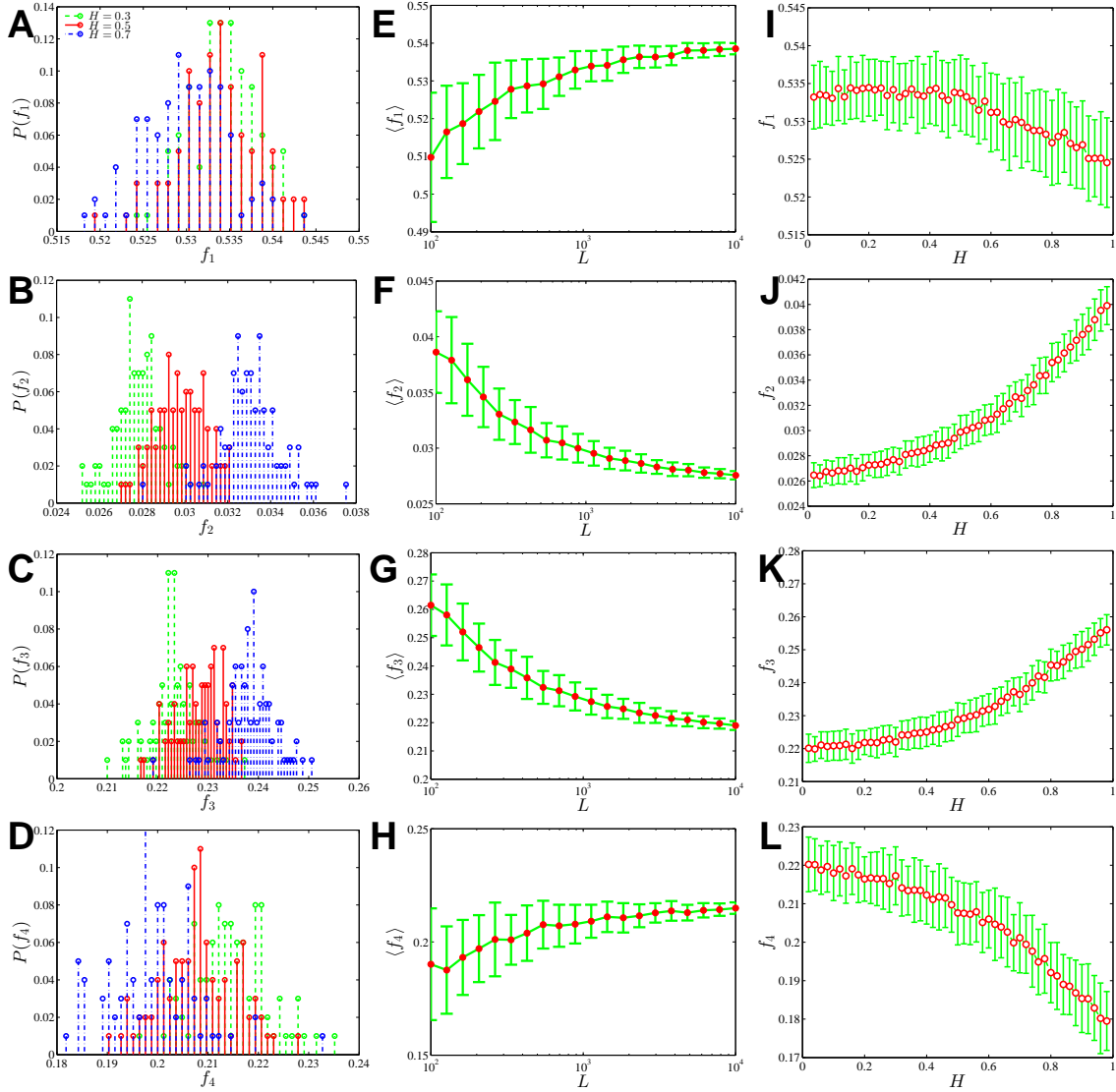}
  \caption{(Color online.) (A-D) Distribution of the occurrence frequencies $f_i$ of motifs $M_i$ for fractional Gaussian noises with Hurst exponent $H=0.3$, $H=0.5$, $H=0.7$.  (E-H) Dependence of the occurrence frequencies $f_i$ of motifs on the length $L$ of time series with Hurst exponent $H=0.5$ for fractional Gaussian noises. The blue and green lines represent the average and the standard deviation of $f_i$ respectively.  (I-L) Dependence of the occurrence frequencies $f_i$ of motifs on the Hurst exponents $H$ for fractional Gaussian noises with length $L$ = 1000. The red dots stands for the average occurrence frequencies and the dashed green lines show the correspond standard deviations.}
\label{Fig:HVG:Mimic:fM:pdf}
\end{figure}

\section{Tetradic motif profiles for fractional Gaussian noises}

To investigate the tetradic motif profiles for fractional Gaussian noises, we conduct numerical simulations. For each Hurst exponent, we generate 100 time series of length $L=1000$. Each time series is transformed into an HVG and the motifs are identified. We note that $N_5=N_6=0$ for fractional Gaussian noises, which can be explained by a similar argument for fractional Brownian motions \cite{Xie-Zhou-2011-PA}.

Fig.~\ref{Fig:HVG:Mimic:fM:pdf} (A-D) show the distribution of the occurrence frequencies $f_i$ for fractional Gaussian noises with different Hurst exponents $H$.  The lines in green, red and blue correspond respectively to $H=0.3$, $H=0.5$, and $H=0.7$. We observe that fractional Gaussian noises with different $H$ have different tetradic motif profiles. This phenomenon is more pronounced for motifs $M_2$ and $M_3$. Hence, tetradic motif profiles have the potential ability to distinguish fractional Gaussian noises with different correlation structures.

We investigate the impact of time series length $L$ on the occurrence frequencies $f_i$ of the motifs. In our simulations, the length $L$ varies from 100 to 10000.
For each length $L$, we generate 100 fractional Gaussian noises with $H=0.5$.
The average occurrence frequencies of motifs for different length $L$ are given in Fig.~\ref{Fig:HVG:Mimic:fM:pdf:L} (E-H). The blue and green lines represent the average and the standard deviation of the occurrence frequencies of motifs respectively.
When the time series length is 10000, the average occurrence frequencies of motifs are close to $(0.5385,    0.0276,    0.2190,    0.2150, 0, 0)$. We find that $f_1$ and $f_4$ increase with $H$, while $f_2$ and $f_3$ decrease with $H$. These tendencies also hold for other values of Hurst exponent.

We further investigate the impact of Hurst exponent $H$ on the occurrence frequencies $f_i$ of motifs for fractional Gaussian noises. In our simulations, the length of time series is fixed to $L$ = 1000. The Hurst exponent $H$ varies from 0.02 to 0.98 with a spacing of 0.02. For each $H$, we simulate 1000 times. Fig.~\ref{Fig:HVG:Mimic:fM:pdf} (I-L) show the results, where the red dots stands for the average occurrence frequencies and the dashed green lines show the correspond standard deviations.
We find that $f_2$ and $f_3$ increase with $H$, while $f_1$ and $f_4$ decrease with $H$.

These findings suggest that the profiles of tetradic HVG motifs depend not only on the correlation structures of time series but also their lengths, showing potential ability to distinguish different time series. In doing so, one has to work on time series with the same length. Otherwise, the distinguished results might be spurious, because the results could be caused by different time series lengths.

\section{Distribution of undirected tetradic motifs for human inter-heartbeat intervals}

We apply the tetradic motif analysis to the heartbeat interval time series and unveil the characteristics and dynamics of different type of heart rate time series. From the PhysioNet website, we retrieve a filtered data set without outliers. These heartbeat interval time series were collected from five healthy subjects, five patients with congestive heart failure (CHF), and five patients with atrial fibrillation (AF). For these 15 subjects, the time series length $N$ varies between 70,000 and 150,000. We divide the time series into sub-time series with length $L$. Therefore each subject's time series is divided into $ \lfloor N/L \rfloor $ non-overlapping sub-series, where we use $L=1000$. For the $ \lfloor N/L \rfloor $ sub-series, we calculate the occurrence frequencies of the five tetradic motifs (motif $M_6$ does not appear).

\begin{figure}[t!]
\centering
\includegraphics[width=14cm]{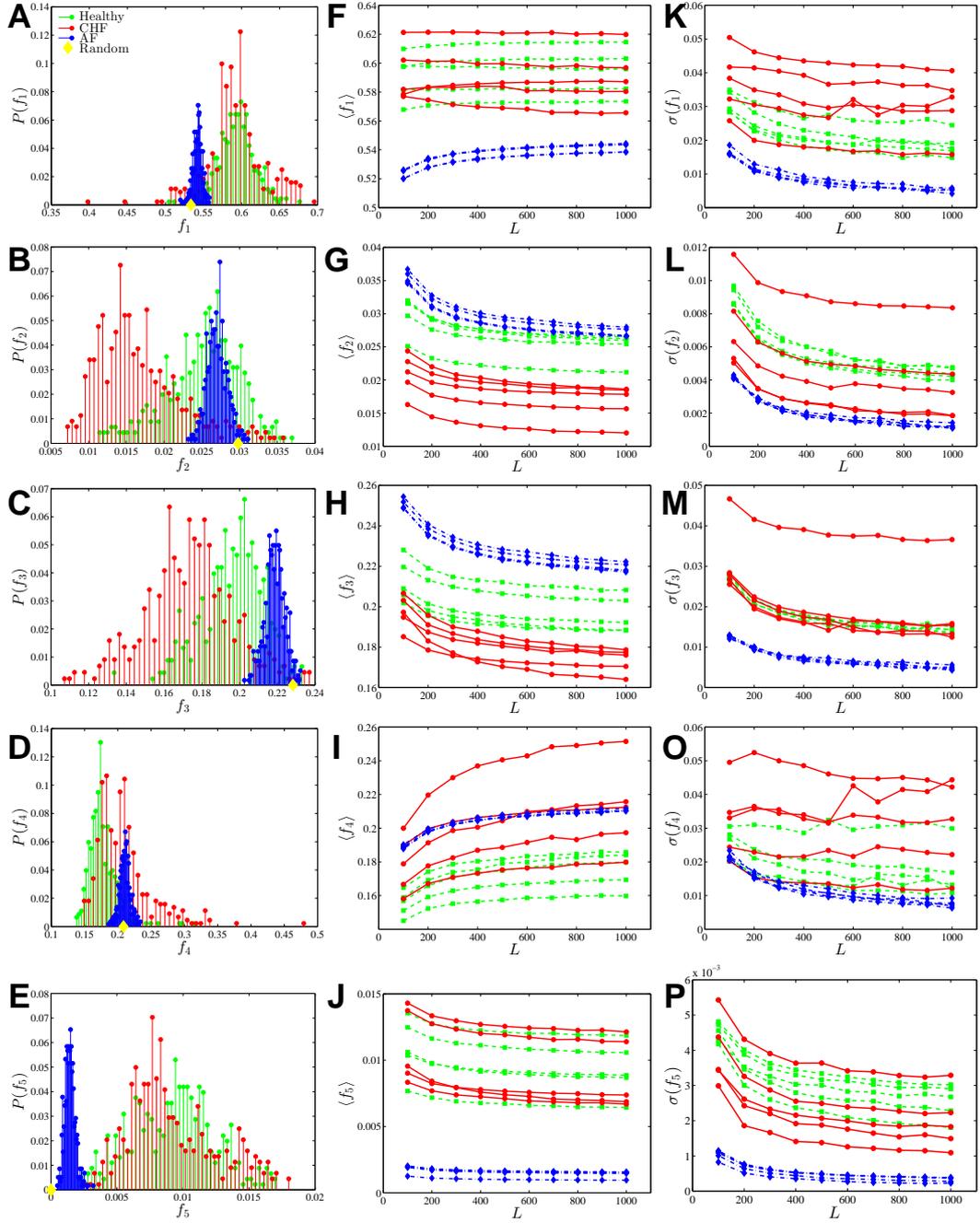}
  \caption{(Color online.) (A-E) Distributions $P(f_{i})$ of occurrence frequency $f_{i}$ of motifs identified from five inter-heartbeat interval time series of healthy subjects (green), CHF patients (red), and AF patients (blue). Each time series of length $N$ is partitioned into $ \lfloor N/L\rfloor $ non-overlapping sub-series of length $L=1000$. We also draw the six values $(0.5334,    0.0298,    0.2283,    0.2085, 0, 0)$ of the uncorrelated time series as filled yellow rhombi for comparison.  (F-J) The average $f_{i}$ in $ \lfloor N/L\rfloor $ sub-series of lengths $L$. Each line corresponds to an individual. The red, green and blue colors are normal, CHF and AF patients.  (K-P) The standard deviation of $f_{i}$.}
\label{Fig:HVG:Mimic:fM:pdf:L}
\end{figure}

For each type of subjects, we put the occurrence frequencies of the motifs of the five subjects together and draw the distribution $P(f_i)$. The $3\times5 = 15$ distributions are shown in Fig.~\ref{Fig:HVG:Mimic:fM:pdf:L} (A-E). In Fig.~\ref{Fig:HVG:Mimic:fM:pdf:L} (A-E), we also draw the five values of uncorrelated time series as filled yellow diamonds for comparison. When the time series length is 1000, the occurrence frequencies of motifs of uncorrelated time series are close to $(0.5334,    0.0298,    0.2283,    0.2085, 0, 0)$. It is found that both the CHF and AF subjects have different motif distributions from the healthy subjects, either in the distribution broadness or in the average values. In addition, the average values for the AF subjects are closer to the uncorrelated case than the healthy and CHF subjects.

We further investigate the impact of time series length on the motif occurrence distributions by looking at the averages and standard deviations of the distributions. The length $L$ off time series varies from 100 to 1000 with a spacing of 100.
For each subject, we determine the occurrence frequencies $f_i$ of motifs in the $ \lfloor N/L \rfloor $ non-overlapping sub-series and calculate the average $\langle{f_i}\rangle$ and standard deviation $\sigma(f_i)$ of the occurrence frequencies of motifs.
Fig.~\ref{Fig:HVG:Mimic:fM:pdf:L} (F-J) show the dependence of the average motif occurrence on the length $L$ of the time series, while Fig.~\ref{Fig:HVG:Mimic:fM:pdf:L} (K-P) show the dependence of the standard deviation of motif occurrences on $L$. In these plots, each line corresponds to a subject, and the colors red, green and blue correspond respectively to the healthy, CHF and AF patients.

We observe in Fig.~\ref{Fig:HVG:Mimic:fM:pdf:L} (F-J) that, except for $\langle{f_1}\rangle$ of the CHF subject, both $\langle{f_1}\rangle$ and $\langle{f_4}\rangle$ increase with $L$ for all subjects, while $\langle{f_2}\rangle$, $\langle{f_3}\rangle$ and $\langle{f_5}\rangle$ decrease with $L$ for all subjects. The $\langle{f_i}\rangle$ curves for motifs $M_2$ and $M_3$ are well separated for the three types of subjects.

In contrast, we observe in Fig.~\ref{Fig:HVG:Mimic:fM:pdf:L} (K-P) that the standard deviation $\sigma(f_i)$ decreases with $L$ for all motifs. In addition, the standard deviation of the AF patients is the smallest. The AF patients can be easily identified according to the standard deviation $\sigma(f_i)$, but not for the healthy and CHF subjects.

\section{Conclusion}

Network motif analysis can identify the characteristics of different time series from different systems.
We use the occurrence frequency of the six undirected tetradic motifs to represent the microstructure features of the HVGs mapped from different time series.
The analytical expression of the motif occurrence frequencies for multifractal binomial measures is obtained, which is consistent with numerical simulations.
We find nonlinear dependence of motif occurrence frequencies on the Hurst exponent and the length of time series for fractional Gaussian noises.
Applying tetradic motif analysis to heartbeat rates, we found that the tetradic motif profiles can distinguish healthy subjects, congestive heart failure subjects, and atrial fibrillation subjects.

In a general sense, we propose to identify different classes of time series through tetradic motif analysis of the horizontal visibility graphs converted from the time series. It requires that the time series must have the same length because time series length affects the HVG motif profiles. We argue that the method should be tested by applying to more time series in other complex systems. It would also be necessary to design more complicated methods to have better identification performance. Moreover, tetradic HVG motifs may contain other useful temporal information of the time series, such as different market states of financial markets of different physical states of sick people. These are meaningful problems deserving further research.

\section*{Acknowledgements}

This work was supported by the National Natural Science Foundation of China [11505063] and Fundamental Research Funds for the Central Universities [222201818006].

%\bibliography{E:/papers/Auxiliary/Bibliography}

%\end{CJK*}
\end{document}